\documentclass[fleqn,twoside]{article}
\usepackage{espcrc2}
\usepackage{epsfig}
\usepackage{graphicx}

\title  {Tunneling conductance in normal metal - triplet superconductor
	junction}

\author{ N. Stefanakis
\address{ Department of Physics, University of Crete,
	P.O. Box 2208, GR-71003, Heraklion, Crete, Greece}
\thanks{Email address: stefan@iesl.forth.gr}}

\begin{document}
 
\begin{abstract}

We calculate the tunneling conductance spectra of a 
normal metal / insulator / triplet superconductor 
from the reflection amplitudes using the 
Blonder-Tinkham-Klapwijk (BTK) formula. 
For the triplet superconductor we assume one special 
$p$-wave order parameter having line nodes and two 
two dimensional $f$-wave order parameters with line 
nodes breaking the time-reversal symmetry. 
Also we examine nodeless pairing potentials. 
The tunneling 
peaks are due to the formation of bound states for
each surface 
orientation at discrete quasiparticles trajectory angles.
The tunneling spectra can be used to 
distinguish the possible candidate pairing states of 
the superconductor Sr$_2$RuO$_4$.
\vspace{1pc}

\end{abstract}

\maketitle

\section{Introduction}
The resent discovery of superconductivity in Sr$_2$RuO$_4$ 
has attracted much theoretical and experimental interest  
\cite{maeno}.
Muon spin rotation experiments show that the time-reversal 
symmetry is broken for the superconductor Sr$_2$RuO$_4$ \cite{luke}.
Knight-shift measurements show no change when passing 
through the superconducting state and is a clear 
evidence for spin triplet pairing state \cite{ishida}.
Also specific heat measurements support the scenario of 
line nodes within the 
gap as in the high $T_c$ cuprate superconductors \cite{nishizaki}.

In tunneling experiments involving singlet superconductors
both line nodes and time-reversal
symmetry breaking can be detected from the V-like shape of the
spectra and the splitting of the zero energy conductance peak (ZEP)
at low temperatures respectively
\cite{blonder,andreev,stefan,stefan1}

Also the properties of ferromagnet - insulator - superconductor with 
triplet pairing symmetry
have been analysed \cite{stefan2}. It is found that the bound states 
are suppressed 
and hence the tunneling conductance peaks 
are eliminated with the increase of the exchange field. 

In this paper we will use the Bogoliubov-de
Gennes (BdG) equations to calculate the
tunneling conductance of normal metal /
triplet superconductor contacts, with a barrier of arbitrary strength
between them, in terms of the
probability amplitudes of Andreev and normal
reflection.
For the triplet superconductor we shall assume
three possible pairing states of two
dimensional order parameter,
having line nodes within the RuO$_2$ plane, which
break the time-reversal symmetry.
The first two are the $2D$ $f$-wave states
proposed by Hasegawa et al,
\cite{hasegawa} having $B_{1g}\times E_u$ and $B_{2g}\times E_u$ symmetry
respectively.
The other one is called nodal $p$-wave state and has been
proposed by Dahm et al \cite{dahm}. This pairing
symmetry has nodes as in the $B_{2g}\times E_u$ case.
Also we will consider the nodeless $p$-wave pairing state 
proposed by
K. Miyake and O. Narikiyo \cite{miyake}.

\section{Theory for the tunneling effect}
\begin{figure}
\psfig{figure=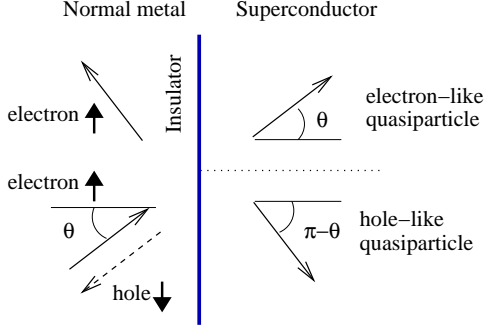,width=6.5cm}
  \caption{
The geometry of the normal metal / insulator / superconductor 
interface. The vertical line along the $y$-axis represents the 
insulator. 
The arrows illustrate the transmission and reflection processes at the 
interface. 
$\theta$ is the angle of the incident electron beam and the normal.
}
  \label{fig1.fig}
\end{figure}

We consider the normal metal  / insulator / superconductor 
junction shown in Fig. \ref{fig1.fig}.
The geometry of the problem has the following limitations.
The particles move in the $xy$-plane
and the boundary between the normal metal
($x<0$) and 
superconductor ($x>0$) is the $yz$-plane at $x=0$. 
The insulator is modeled by a delta function, located at $x=0$, of the 
form $V\delta(x)$. The temperature is fixed to $0$ K.
We take the pair potential
as a step function i.e.
$\Delta_{s\overline{s}}({\bf k},{\bf r})=
\Theta(x)\Delta_{s\overline{s}}(\theta)$, 
where $k_x,k_y=\cos\theta, \sin\theta$, and $s,\overline{s}$ are 
spin indices.

Suppose that an electron is incident from the normal metal
to the insulator with an angle $\theta$. The electron like 
(hole) like quasiparticle will experience different pair potentials 
$\Delta_{s\overline{s}}(\theta)(\Delta_{s\overline{s}}(\pi-\theta))$.
The amplitudes for Andreev and normal reflection are obtained 
by solving the BdG equations.
Using the matching conditions of the wave function at $x=0$,
$\Psi_I(0)=\Psi_{II}(0)$ 
and 
$\Psi_{II}'(0)-\Psi_{I}'(0)=(2mV/\hbar^2)\Psi_I(0)$,
the Andreev and normal reflection amplitudes
$R_a=|a|^2,R_b=|b|^2$
are obtained

\begin{equation}
R_a=\frac{\sigma_N^2|n_{+}|^2}
     {|1+(\sigma_N-1)n_{+}n_{-}
      \phi_{-}\phi_{+}^{\ast}|^2}
,~~~\label{ra}
\end{equation}

\begin{equation}
R_b=\frac{(1-\sigma_N)|1-n_{+}n_{-} 
     \phi_{-}\phi_{+}^{\ast}|^2}
     {|1+(\sigma_N-1)n_{+}n_{-}
     \phi_{-}\phi_{+}^{\ast}|^2}
.~~~\label{rb}
\end{equation}
The BCS coherence factors are given by
\begin{equation}
u_{\pm}^2=[1+
      \sqrt{E^2-|\Delta_{\pm}(\theta)|^2}/E]/2,
\end{equation}

\begin{equation}
v_{\pm}^2=[1-
      \sqrt{E^2-|\Delta_{\pm}(\theta)|^2}/E]/2,
\end{equation}
and $n_{\pm}=v_{\pm}/u_{\pm}$.
The internal phase coming from the energy gap is given by
$\phi_{\pm} =[
\Delta_{\pm}(\theta)/|\Delta_{\pm}(\theta)|]$,
where $\Delta_{+}(\theta)=\Delta(\theta)$
($\Delta_{\_}(\theta)=\Delta(\pi- \theta)$), is the 
pair potential experienced by the transmitted electron-like 
(hole-like) quasiparticle.
$\Delta(\theta)=\Delta_{\uparrow \downarrow}(\theta)$
$=\Delta_{\downarrow \uparrow}(\theta)$, since the Cooper 
pairs have zero spin projection i.e. ${\bf d} \parallel \hat {\bf z}$.

The tunneling conductance, normalized by that in the normal 
state is given by \cite{blonder}
\begin{equation}
\sigma(E)=\frac{\int_{-\pi/2}^{\pi/2}d\theta
(\overline{\sigma}_{\uparrow}(E,\theta)+
 \overline{\sigma}_{\downarrow}(E,\theta))}
{\int_{-\pi/2}^{\pi/2}d\theta2\sigma_N }
.~~~\label{ss}
\end{equation}
According to the BTK formula the conductance of the junction 
$\overline{\sigma}_s(E,\theta)$, for spin 
$s=\uparrow,\downarrow$,  
is expressed in terms of the 
probability amplitudes $a$, and $b$ as
\begin{equation}
\overline{\sigma}_s(E,\theta)=1+R_a-R_b. 
\end{equation}
The transparency of the junction $\sigma_N$ is connected to 
the barrier height $V$ by the relation 
\begin{equation}
\sigma_N=\frac{4 \cos^2\theta}{Z^2+4 \cos^2\theta}
,~~~\label{sn}
\end{equation}
where $Z=2 m V / \hbar^2 k_F$, denotes the strength of the barrier. 

The pairing potential is described by a $2\times 2$ form
\begin{equation}
\hat{\Delta}_{s\overline{s}}({\bf k}) =
  \left(
    \begin{array}{ll}
     -d_x({\bf k}) +id_y({\bf k})&
     d_z({\bf k}) \\
     d_z({\bf k}) &
    d_x({\bf k}) +id_y({\bf k})
    \end{array}
  \right)
,~~~\label{deltafour}
\end{equation}
in terms of the $d({\bf k})=(d_x({\bf k}),d_y({\bf k}),d_z({\bf k}))$ vector.

We consider the following pairing symmetries for Sr$_2$RuO$_4$.
 
a) In the first $2D$ $f$-wave state $B_{1g}\times E_u$
$d_z({\bf k})=\Delta_0(k_x^2-k_y^2)(k_x+ik_y)$.
This state has nodes at the same points as in the $d_{x^2-y^2}$-wave
case.

b) For the second $2D$ $f$-wave state $B_{2g}\times E_u$
$d_z({\bf k})=\Delta_0k_xk_y(k_x+ik_y)$.
This state has nodes at $0, \pi/2, \pi, 3\pi/2$ and has also 
been studied by Graf and Balatsky \cite{graf}.

c) In case of a nodal $p$-wave superconductor
$d_z({\bf k})=
\frac{\Delta_0}{s_M}[\sin(k_xa)+i\sin(k_ya)]$, with 
$k_xa=R \pi \cos(\theta-\beta)$ and $k_ya=R \pi \sin(\theta-\beta)$,
$s_M=\sqrt{2}\sin\frac{\pi}{\sqrt{2}}=1.125$, and $R=1$ in order to have
a node in $\Delta(\theta)$ \cite{dahm}.
This state has nodes as in the $B_{2g}\times E_u$ state.

d) In case of a nodeless $p$-wave superconductor, proposed by K. Miyake, and O. Narikiyo
\cite{miyake}
$d_z({\bf k})=
\frac{\Delta_0}{s_M}[\sin(k_xa)+i\sin(k_ya)]$,
with $k_xa=R \pi \cos(\theta-\beta)$ and $k_ya=R \pi \sin(\theta-\beta)$,
$s_M=\sqrt{2}\sin\frac{\pi}{\sqrt{2}}=1.125$, and $R=0.9$.
This state does not have nodes.

\section{Results}
\begin{figure}
  \psfig{figure=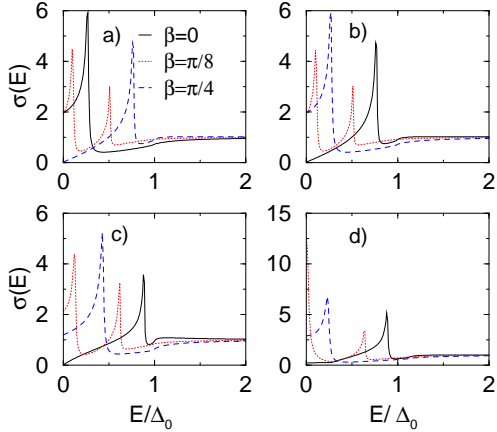,width=6.5cm}
  \caption{
Normalized tunneling conductance $\sigma(E)$ as a function of $E/\Delta_0$
for different orientations
$\beta=0$ (solid line), $\pi/8$ (dotted line), $\pi/4$ (dashed line).
The temperature is $T=0$, $Z=10$.
The pairing 
symmetry of the superconductor is 
(a) $B_{1g}\times E_u$, (b) $B_{2g}\times E_u$, (c) nodal $p$-wave,
(d) nodeless $p$-wave.
}
  \label{Z10.fig}
\end{figure}

In Fig. \ref{Z10.fig}
we plot the tunneling conductance $\sigma(E)$ 
as a function of $E/ \Delta_0$
for $Z=10$, for different orientations (a) $\beta=0$, 
(b) $\pi/8$, 
(c) $\pi/4$. The pairing 
symmetry of the superconductor is 
$B_{1g}\times E_u$ in Fig. \ref{Z10.fig}a, 
$B_{2g}\times E_u$ in Fig. \ref{Z10.fig}b, 
nodal $p$-wave, in Fig. \ref{Z10.fig}c, 
nodeless $p$-wave in Fig. \ref{Z10.fig}d.
The temperature is fixed to $0K$.
In all the cases we see the presence of a large residual 
density of states within the energy gap and peaks due to the sign change 
of the pair potential at discreet values
of $\theta$, for fixed $\beta$. 
Also the linear increase with $E$ 
is consistent with the presence on line nodes in the 
pairing potential. 

Generally the spectra  for the three pairing symmetries 
with line nodes depends strongly on the position of the 
nodes in the pairing potential
and the orientation angle $\beta$. 
The spectra for angle $\beta$ in the $B_{1g}\times E_u$ 
seen in Fig. \ref{Z10.fig}a
case is identical to the spectra of $B_{2g}\times E_u$ in Fig. \ref{Z10.fig}b
for 
angle $\pi /4-\beta$, since the node positions for the two symmetries 
differ by $\pi /4$.
The nodal $p$-wave case in Fig.  \ref{Z10.fig}c
has the same nodal structure as the 
$B_{2g}\times E_u$ case and we see that the spectra for these two 
candidates are similar. 
For nodeless pairing states a subdap or a full gap opens in the tunneling 
spectra.
This is seen in Fig. \ref{Z10.fig}d, 
for the nodeless $p$-wave pairing state.

\begin{figure}
  \psfig{figure=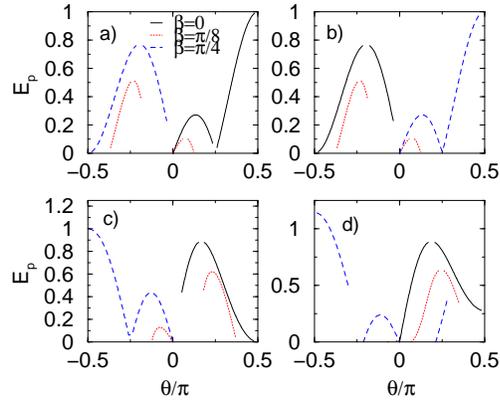,width=6.5cm}
  \caption{
Bound state energy $E_p$ (in units of $\Delta_0$) for $\beta=0, \pi /8,\pi /4$
as a function of $\theta$. The temperature is $T=0$.
The pairing 
symmetry of the superconductor is 
(a) $B_{1g}\times E_u$, 
(b) $B_{2g}\times E_u$,
(c) nodal $p$-wave,
(d) nodeless $p$-wave.
}
  \label{bs.fig}
\end{figure}
The peaks in the tunneling conductance are explained from the 
formation of bound states within the gap.
The bound state energies $E_p$,
are given from the values of $E$ where the
denominator of Eqs. (\ref{ra},\ref{rb}) vanishes. In this case
the Andreev reflection coefficient is equal to unit, and the
effect of the boundary is turned off.
The corresponding equation is written as  \cite{stefan}
\begin{equation}
     \phi_{-}\phi_{+}^{\ast}n_{+}n_{-}|_{E=E_p}=1.0
      .~~~\label{midgap}
\end{equation}
Bound states are formed because the transmitted electron-like
and hole-like quasiparticles feel different sign of the
pairing potential.
These bound states occur for a given 
orientation $\beta$ at discreet values of $\theta$, 
as seen in Fig. \ref{bs.fig} where the 
bound state energy $E_p$ is plotted for $\beta=0, \pi /8,\pi /4$ 
as a function of $\theta$ for the various pairing states.
For a given value of $\beta$, 
the conductance  $\overline \sigma (E,\theta)=2$
at the angles $\theta$ where bound state occurs 
and 
the tunneling conductance $\sigma(E)$ is enhanced due to 
the normal state conductance $\sigma_N$. 
The residual values of the tunneling conductance within the gap 
is due to the 
bound states and the peaks are formed at energies where the 
number of bound states is increased. 
As seen in Fig. \ref{bs.fig} (a) for the pairing state $B_{1g} \times E_u$
for $\beta=0$, at the energy in the interval $0.27<E<1$ only 
one bound state occurs. At $E=0.27$ two more bound states are 
formed and the peak seen in Fig. \ref{Z10.fig} (a) is due to these 
new bound states. 
The same happens for $\beta=\pi /8$ at $E=0.1$
and $E=0.5$, where an increased number of bound states is formed,
as seen in Fig \ref{Z10.fig} (a) (dotted line).
Also for $\beta=\pi /4$ in Fig \ref{bs.fig} (a) (dashed line)
the conductance peak is formed for energy close to $E=0.8$, where 
two bound states are formed.
In the $B_{2g} \times E_u$ pairing state the bound states and also
the position of the peaks for an angle $\beta$ is the same as in the
$B_{1g} \times E_u$ pairing state for the angle $\pi /4-\beta$.
In the nodal $p$-wave state seen in Fig. \ref{bs.fig} (c)
the bound states are symmetric to the
$B_{2g} \times E_u$ case with respect to $\theta=0$.
However this does not influences
the energy levels which occur almost at the same position as in the
$B_{2g} \times E_u$-wave case.
In the nodeless $p$-wave pairing state,
for $\beta=\pi /8$ close to $E=0$ we have a pair of new bound states
and the peak seen in Fig. \ref{Z10.fig} (d) is due to these new bound states.

We calculated the tunneling conductance in normal metal / insulator / 
triplet superconductor using the BTK formalism. We assumed nodal and 
nodeless
pairing potentials, 
breaking the time-reversal 
symmetry.
The residual values are due to the formation of bound states 
for each $\beta$ at discreet values of the angle $\theta$ for 
energies within the gap. The peaks occurs at the energies where 
an increasing number of bound states is formed.

\end{document}